\documentclass[conference]{IEEEtran}
\IEEEoverridecommandlockouts

\usepackage{cite}
\usepackage{amsmath,amssymb,amsfonts}
\usepackage{algorithmic}
\usepackage{graphicx}
\usepackage{textcomp}
\usepackage{xcolor}
\usepackage[hyphens]{url}
\usepackage{multirow}
\usepackage{makecell}
\usepackage{booktabs}
\usepackage[hidelinks]{hyperref}
\usepackage{authblk}

\usepackage{xcolor}

\def\BibTeX{{\rm B\kern-.05em{\sc i\kern-.025em b}\kern-.08em
    T\kern-.1667em\lower.7ex\hbox{E}\kern-.125emX}}

\begin{document}

\pdfpagewidth=8.5in
\pdfpageheight=11in


\pagenumbering{arabic}

\title{Bridging Superconducting and Neutral-Atom Platforms for Efficient Fault-Tolerant Quantum Architectures}

\author[1]{Xiang Fang}
\author[1]{Jixuan Ruan}
\author[1]{Sharanya Prabhu}
\author[2]{Ang Li}
\author[3]{Travis Humble}
\author[1]{Dean Tullsen}
\author[1]{Yufei Ding}

\affil[1]{University of California, San Diego, CA, USA}
\affil[2]{Pacific Northwest National Laboratory, Richland, WA, USA \& University of Washington, Seattle, WA, USA}
\affil[3]{Quantum Science Center, Oak Ridge National Laboratory, Oak Ridge, TN, USA}

\date{}

\maketitle
\thispagestyle{plain}
\pagestyle{plain}

\begin{abstract}
The transition to the fault-tolerant era exposes the limitations of homogeneous quantum systems, where no single qubit modality simultaneously offers optimal operation speed, connectivity, and scalability. In this work, we propose a strategic approach to Heterogeneous Quantum Architectures (HQA) that synthesizes the distinct advantages of the superconducting (SC) and neutral atom (NA) platforms. We explore two architectural role assignment strategies based on hardware characteristics: (1) We offload the latency-critical Magic State Factory (MSF) to fast SC devices while performing computation on scalable NA arrays, a design we term MagicAcc, which effectively mitigates the resource-preparation bottleneck. (2) We explore a Memory-Compute Separation (MCSep) paradigm that utilizes NA arrays for high-density qLDPC memory storage and SC devices for fast surface-code processing. Our evaluation, based on a comprehensive end-to-end cost model, demonstrates that principled heterogeneity yields significant performance gains. Specifically, our designs achieve $752\times$ speedup over NA-only baselines on average and reduce the physical qubit footprint by over $10\times$ compared to SC-only systems. These results chart a clear pathway for leveraging cross-modality interconnects to optimize the space-time efficiency of future fault-tolerant quantum computers.
\end{abstract}
\section{Introduction} \label{sec: intro}

\begin{figure*}[!ht]
    \centering
    \includegraphics[width=.99\textwidth]{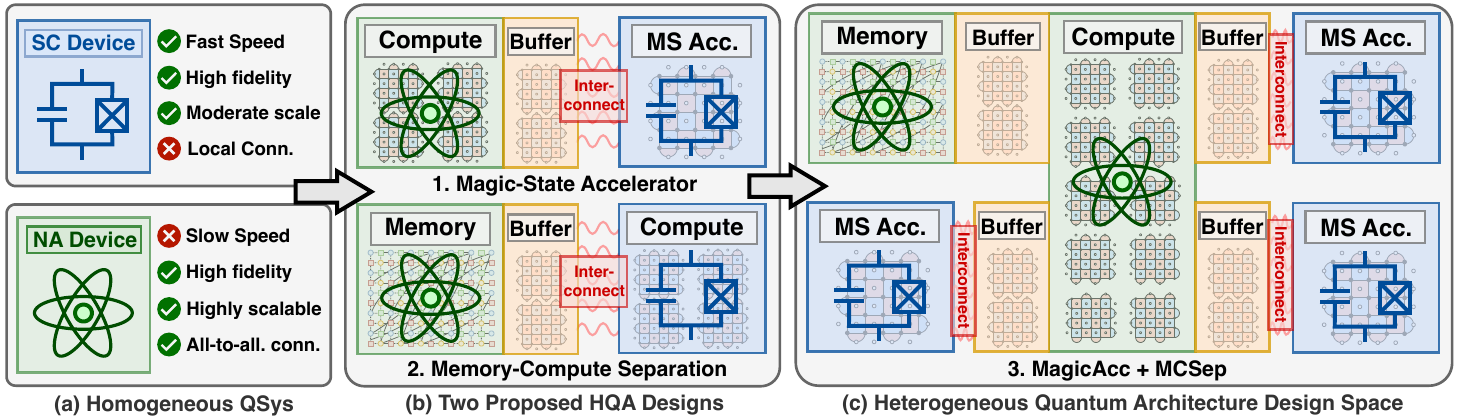}
    \vspace*{-6pt}
    \caption{Overview of proposed heterogeneous quantum architectures and the derived design space.}
    \vspace*{-10pt}
    \label{fig: introduction}
\end{figure*}

The rapid evolution of quantum hardware across superconducting (SC)~\cite{arute2019quantum, bravyi2022future, google2023suppressing, natureReleasesFirstever, acharya2024quantumerrorcorrectionsurface}, trapped ion (TI)~\cite{bruzewicz2019trapped, strohm2024ion, moses2023race, ransford2025helios98qubittrappedionquantum}, neutral atom (NA)~\cite{bluvstein2024logical, bluvstein2025architectural, wintersperger2023neutral}, and emerging platforms~\cite{psiquantum2025manufacturable, zhang2025demonstration, pezzagna2021quantum} brings fault-tolerant quantum computing (FTQC)~\cite{shor1996fault} with quantum error correction (QEC)~\cite{gottesman1997stabilizer} closer to reality. This progress promises to unlock transformative applications in simulation and optimization~\cite{gottesman1998theory, shor1996fault, shor1999polynomial, cao2019quantum, daley2022practical, bauer2020quantum, pearson2020simulating, liu2024toward, alexeev2025perspective}. Yet, a fundamental barrier remains: no single physical modality simultaneously offers fast gate speeds, high fidelity, long coherence, flexible connectivity, and mass manufacturability. This limitation constrains the scalability of homogeneous systems. Concurrently, breakthroughs in quantum interconnects and transduction~\cite{lauk2020perspectives, wang2022quantum, caleffi2025quantum, fan2018superconducting, han2020cavity, mirhosseini2020superconducting, darpa2025harq, dhordjevic2021entanglement, beukers2024remote, huang2022two, zhang2024entanglement} are making cross-platform integration increasingly viable, paving the way for \emph{heterogeneous quantum architectures} (HQA) that leverage the distinct strengths of multiple modalities to achieve system-level advantages.

Heterogeneity in classical computing emerged from the necessity of specialization—utilizing CPUs for control~\cite{hennessy2011computer}, GPUs for data parallelism~\cite{nickolls2008scalable}, and AI accelerators for tensor operations~\cite{jouppi2017datacenter}. We apply this architectural principle to the quantum domain. Instead of treating different modalities as generic compute units for partitioned sub-circuits~\cite{tang2021cutqc, ren2024hardware}, we advocate for assigning \textit{system roles} based on the alignment between hardware characteristics and workload demands. This strategic assignment is essential for realizing tangible gains over homogeneous designs.

Realizing this vision, however, requires overcoming three distinct challenges: \textbf{(C1) Modality Mismatch.} Hardware characteristics—speed, fidelity, scalability, and connectivity—vary drastically between platforms. Without careful role assignment, the performance limitations of one device can bottleneck the entire system, negating heterogeneous benefits. \textbf{(C2) Resource-Prep Bottleneck.} FTQC stacks involve not just computation but also resource preparation pipelines (e.g., magic state distillation for non-Clifford gates). These processes often dominate the space–time budget and introduce stochastic delays~\cite{litinski2019magic, gidney2024magic}. \textbf{(C3) Interconnect Constraints.} Despite advances, interconnects impose realistic constraints on bandwidth, latency, and loss. Architectures must ensure logical correctness while minimizing transport-induced stalls. Successful HQA designs must therefore anticipate whether bottlenecks lie in compute, resource preparation, or transport.

In this work, we select the SC–NA pairing as a primary case study to address these challenges. These two modalities exhibit complementary profiles (Fig.\ref{fig: introduction}(a)), exemplifying challenge (C1). SC platforms provide fast ($10$–$100\, ns$), high-fidelity ($99.9\%$) gates but are limited to moderate scales ($10^2$–$10^3$ qubits)~\cite{arute2019quantum, google2023suppressing, acharya2024quantumerrorcorrectionsurface, natureReleasesFirstever} with fixed, local connectivity. Conversely, NA platforms~\cite{bluvstein2024logical, bluvstein2025architectural, levine2022dispersive, evered2023high} feature massive scalability (over 6{,}000 qubits~\cite{manetsch2024tweezer, chiu2025continuous}) and reconfigurable all-to-all connectivity, but operate with significantly slower gate times ($0.1$–$1\,\mu s$) and transport speeds ($0.1$–$1\, ms$). HQA designs must therefore reconcile mismatches in \emph{connectivity} (flexible NA vs. local SC), \emph{scalability} (NA advantage), and \emph{speed} (SC advantage).

To resolve these disparities and address challenges (C1)-(C3), we propose two distinct HQA design methodologies:

\noindent \textbf{(1) Magic state Accelerator (MAcc)}. This methodology targets the FTQC cost asymmetry between resource-intensive magic state factories (MSF) and efficient Clifford operations~\cite{bravyi2005universal}. We exploit the fact that MSF requires high volume but only local connectivity, whereas Clifford layers benefit from flexible connectivity for transversal execution~\cite{cain2024correlated, zhou2025low, zhou2025resource, serra2025decoding}. The $\sim10^3$x cost difference between these operations mirrors the $\sim10^3$x speed difference between SC and NA. MAcc (Fig.\ref{fig: introduction}(b1)) strategically assigns the MSF workload to fast SC devices to accelerate supply, while mapping Clifford compute to scalable NA arrays. This effectively balances latency (C1) and alleviates the resource-preparation bottleneck (C2). Furthermore, since magic states are distinct from program data, their transport decouples preparation from execution, mitigating interconnect risks (C3).

\noindent \textbf{(2) Memory–Compute Separation (MCSep).} Drawing on the classical hierarchy of CPU and DRAM~\cite{hennessy2011computer}, we propose treating fast SC devices as the \emph{compute engine} and scalable NA arrays as the \emph{memory}. This design addresses the speed mismatch (C1) and leverages connectivity differences to optimize QEC encoding. SC qubits with nearest-neighbor coupling are ideal for surface codes~\cite{bravyi1998quantum, dennis2002topological, litinski2019game, fowler2012surface, beverland2022assessing}, which enable fast logical operations but incur high overhead. In contrast, NA qubits with non-local connectivity support qLDPC codes~\cite{tillich2013quantum, panteleev2022asymptotically, panteleev2021degenerate, bravyi2024high, leverrier2022quantum, dinur2023good, breuckmann2021quantum, bravyi2013classification, wang2024coprime, xu2025batched}, offering orders-of-magnitude space savings~\cite{xu2024constant, yoder2025tour} at the cost of operation complexity. MCSep (Fig.~\ref{fig: introduction}(b2)) dynamically loads logical qubits from NA qLDPC memory into the SC compute region, achieving high execution speed with a minimal spatial footprint. We also explore an intra-modality variant (NA memory + NA compute) to expand the design space.

We integrate MAcc and MCSep into a unified HQA design space (Fig.\ref{fig: introduction}(c)) and introduce a systematic framework for modeling end-to-end runtime and resource costs (Sec.\ref{sec:arch_modeling}). Through architectural analyses (Sec.\ref{sec: arch analysis}), including Amdahl-style speedup formulations and space–time trade-off projections, we link benchmark characteristics (e.g., T-layer ratio, Pauli-weight density) to optimal architectural choices.

Our extensive evaluation demonstrates that principled heterogeneity yields substantial gains: \textbf{MAcc} achieves $500$–$1000\times$ speedup over NA-only baselines, while \textbf{MCSep} reduces physical qubit requirements by up to $10.8\times$ compared to SC-only designs. Design space exploration further reveals how factors such as transport latency and buffer sizing influence these trade-offs, enabling us to distill practical rules for mapping applications to architectures. While grounded in the NA–SC pairing, our framework is generalizable to other modality combinations.

In summary, this paper contributes:
\begin{itemize}
\item A \emph{role-assignment} design principle for heterogeneous FTQC systems that exploits complementary hardware strengths.
\item Two concrete HQA methodologies, \emph{MAcc} and \emph{MCSep}, enabling magic-state acceleration and memory-compute separation respectively.
\item A comprehensive framework for HQA modeling, evaluation, and architectural analysis.
\item A quantitative exploration of the HQA design space, offering insights into trade-offs and application-architecture mapping.
\end{itemize}
\section{Background and Related Work}
\label{sec:background}
We first introduce FT workloads, an abstraction that exposes the FTQC system cost landscape (Sec.~\ref{subsec:ft-workloads}), and review the two execution schemes used on NA and SC (Sec.~\ref{subsec:compute_schemes}). We then introduce the resource-intensive MSF (Sec.~\ref{subsec:MSF}) and the emerging quantum interconnect technologies (Sec.~\ref{subsec:interconnect}).

\subsection{FT Workloads: Architectural Abstraction}
\label{subsec:ft-workloads}

Studies of FTQC systems span several abstraction levels, each exposing different facets of implementation cost.

\noindent\textit{1. Algorithmic circuit.} At the top, the program is a purely functional specification with no QEC or hardware considerations. This level is useful for correctness and asymptotic reasoning but hides all FT overheads~\cite{montanaro2016quantum, dalzell2023quantum, martyn2021grand}.

\noindent\textit{2. Synthesized circuit.} The algorithmic circuit is mapped to a target logical gate set (e.g., Clifford+T~\cite{ross2014optimal}), yielding metrics such as two-qubit count, $T$-count, depth and enabling some level of optimizations~\cite{wang2025tableau, wang2024optimizing, liu2025quclear, GridSynth}. However, this representation is still without QEC integration and thus remains too abstract for realistic FT cost estimation.

\noindent\textit{3. FT workloads.} The next level selects QEC codes and logical operation protocols and reveals ancillary FT routines (e.g., syndrome measurements~\cite{yin2025qecc}, resource-state preparation~\cite{bravyi2005universal}, decoding~\cite{higgott2022pymatching, roffe2020decoding}). This representation begins to expose the cost landscape (e.g., per-layer cycle structure, ancilla pipelines, module interaction) while remaining hardware-agnostic and amenable to parameterized modeling.

\noindent\textit{4. Physical realization.} Finally, FT workloads are mapped and scheduled onto concrete devices, accounting for platform-specific constraints. Numerous toolchains target this level for particular modalities~\cite{li2019tackling, ruan2025powermove, zhang2023oneq, ruan2025trapsimd, wu2022synthesis, zhu2025s, tan2024compiling, molavi2022qubit, zhang2024oneperc}, but the details are highly hardware dependent.

\noindent\textit{Most architecture studies therefore reason at the FT-workload level}~\cite{gidney2021factor, zhou2025low, fowler2018low, litinski2019game, litinski2019magic, beverland2022assessing, zhou2025resource, yoder2025tour}: it exposes the dominant cost drivers without over-committing to device-level minutiae that can obscure system insights. Concrete resource estimation are supplied by parameterized cost models calibrated to a given hardware setting.

\begin{figure}[!ht]
    \centering
    \vspace{-6pt}
    \includegraphics[width=0.48\textwidth]{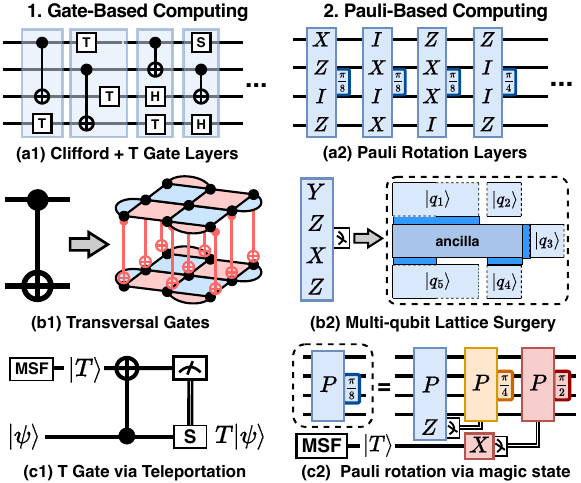}
    \vspace{-6pt}
    \caption{Background on FTQC. (a) GBC and PBC layers. (b) Logical operation on surface codes (transversal gates vs. lattice surgery). (c) Realizing non-Clifford operations by consuming the magic state $|T\rangle = (|0\rangle+e^{i\frac{\pi}{4}}|1\rangle)/\sqrt{2}$).}
    \label{fig:background} 
    \vspace{-6pt}
\end{figure}

\subsection{Execution Schemes: GBC and PBC}
\label{subsec:compute_schemes}

We introduce two leading surface-code FTQC schemes and their natural alignment with NA and SC devices.

\noindent\textbf{Gate-Based Computing (GBC) on NA.}
GBC organizes computation into \emph{gate layers} (Fig.~\ref{fig:background}(a1)). The gates within a layer can execute concurrently. On NA platforms, surface-code Clifford gates (CNOT, H, S) admit efficient \emph{transversal} realization (Fig.~\ref{fig:background}(b1)), enabled by flexible connectivity through atom transport~\cite{bluvstein2024logical}. Non-Clifford $T$ gates are implemented by consuming magic states (see Sec.~\ref{subsec:MSF}). A further advantage on NA is that one round of syndrome measurement (SM) per gate layer typically suffices for fault tolerance~\cite{cain2024correlated, zhou2025low, cain2025fast}, making GBC a leading choice for NA devices.

\noindent\textbf{Pauli-Based Computing (PBC) on SC.}
PBC is based on layers of Pauli-product rotations~\cite{litinski2019game, litinski2019magic, fowler2018low} (Fig.~\ref{fig:background}(a2)): rotation angles $\{\frac{\pi}{2},\frac{\pi}{4}\}$ realize Clifford operations, while $\frac{\pi}{8}$ angle gives non-Clifford operation and consume magic states (see Sec.~\ref{subsec:MSF}). These rotations are realized using Pauli-product measurements (PPMs) via lattice surgery~\cite{vuillot2019code, horsman2012surface, chamberland2022universal}, requiring only local interactions between surface-code boundaries (Fig.~\ref{fig:background}(b2)), well matched to SC devices with local connectivity. Unlike GBC with transversal gates, each PPM incurs $d_{\text{surf}}$ SM rounds that scales as the code distance. Although either scheme can in principle run on either modality, pairing \emph{GBC with NA} and \emph{PBC with SC} better leverages native platform capabilities and is commonly adopted~\cite{beverland2022assessing, gidney2021factor, litinski2019game, zhou2024algorithmic, zhou2025resource}.

\subsection{Non-Clifford operations and Magic State Factory}\label{subsec:MSF}

\noindent\textbf{Realizing non-Clifford gates.}
Non-Clifford gates (e.g., $T$ gates) are essential for universal QC but are difficult to realize directly on most QEC codes due to fundamental limitations~\cite{eastin2009restrictions, bravyi2013classification, fu2025no}. A common workaround is \emph{gate teleportation}~\cite{gottesman1999demonstrating, knill2004fault, zhou2000methodology}, which implements non-Clifford gates using Clifford gates and pre-prepared \emph{magic states}~\cite{bravyi2005universal} (e.g., consume $|T\rangle$ magic for T gate~\cite{bravyi2005universal}, Fig.~\ref{fig:background}(c1)). While Clifford gates are typically cheap, generating high-fidelity $|T\rangle$ states requires a resource-intensive \emph{magic state factory} (MSF). Changing the computing scheme may alter how magic states are consumed but cannot avoid them. For example, in PBC, each $\frac{\pi}{8}$-Pauli rotation still consumes one magic state via PPM~\cite{litinski2019game, fowler2018low} and requires MSF (Fig.~\ref{fig:background}(c2)).

\noindent\textbf{From distillation to cultivation.}
The conventional \emph{magic state distillation} (MSD)~\cite{bravyi2005universal} approach to MSF remains highly resource-intensive despite consistent improvement efforts~\cite{bravyi2012magic, haah2018codes, hastings2018distillation, gidney2019efficient, litinski2019magic}. A representative protocol~\cite{litinski2019magic} requires nearly 5,000 physical qubits to produce $|T\rangle$ states with infidelity $10^{-8}$, beyond the capacity of most individual hardware nodes. Recently, \emph{magic state cultivation} (MSC) has emerged~\cite{gidney2024magic, vaknin2025magic, sahay2025fold}, trading space for time: it uses far fewer physical qubits (e.g., 463 qubits~\cite{gidney2024magic}) and only neighboring connectivity, but relies on postselection. Depending on the target fidelity, each successful output may require 20–100 attempts~\cite{gidney2024magic} costing hundreds or thousands of QEC cycles.

\begin{table*}[t]
\centering
\caption{Heterogeneous Quantum Operation Set (HQ-ISA): categories, instructions, and semantics.}
\label{tab:hqisa}
\small
\begin{tabular}{l l p{8.8cm}}
\toprule
\textbf{Category} & \textbf{Instruction} & \textbf{Semantics and Purpose} \\
\midrule

\multirow{8}{*}{\textbf{\makecell[c]{Compute (NA/SC)}}}
& \texttt{NA\_GBC\_LAYER(spec,r)} & Execute one NA GBC logical layer (transversal Clifford gates followed by \texttt{r} SM rounds) according to implementation specifications \texttt{spec}; consumes magic states if the gate layer contains T gates. \\
& \texttt{SC\_PBC\_PPM(spec,r)} & Execute one SC PPM layer via lattice surgery with $r$ SM rounds according to \texttt{spec}; consumes one magic state per layer. \\
& \texttt{MAGIC\_TAKE(m)} & Consume $m$ magic states from the compute-side buffer; blocks until available (use \texttt{FENCE\_MAGIC} to ensure availability beforehand). \\
& \texttt{FENCE\_MAGIC} & Ordering/visibility fences: ensure prior measurements or magic deliveries are visible before dependent/consuming steps. \\
\midrule

\multirow{3}{*}{\textbf{\makecell[c]{Memory (NA)}}}
& \texttt{LDPC\_STORE(qset)} & Store logical qubits \texttt{qset} into NA qLDPC memory for later retrieval. \\
& \texttt{LDPC\_LOAD(qset)} & Retrieve logical qubits \texttt{qset} from memory into the compute region. \\
& \texttt{BUF\_RESERVE(Q)} & Reserve/size the NA logical buffer capacity $Q$ for inter-layer swaps. \\
\midrule

\multirow{1}{*}{\textbf{\makecell[c]{MSF (NA/SC)}}}
& \texttt{MSF\_START(host,k)} & Activate $k$ parallel MSF instances on the MSF \texttt{host}. \\
\midrule

\multirow{3}{*}{\textbf{\makecell[c]{\\ \\ Transport \\ (NA--SC)}}}
& \texttt{SEND\_MAGIC(host,m)} & Send $m$ magic states \emph{from} MSF instance \texttt{src} to the compute-side magic-state buffer over the NA--SC link (used in MAcc). \\
& \texttt{RECV\_MAGIC(host,m)} & Receive $m$ magic states into the compute-side buffer from \texttt{host}; marks them available for later consumption (used in MAcc). \\
& \texttt{TRANS\_LQ(dir,qset)} & Transfer the logical-qubit set \texttt{qset} across modalities in direction \texttt{dir} $\in$ \{NA--SC, SC--NA\} (used in MCSep). \\
\bottomrule
\end{tabular}
\end{table*}

\subsection{Quantum Transduction and Interconnect}\label{subsec:interconnect}
Quantum transduction~\cite{lauk2020perspectives, wang2022quantum, caleffi2025quantum} refers to the coherent conversion of quantum information between physically distinct carriers, such as from microwave photons in SC systems to optical photons in NA systems. This conversion is the critical technology underpinning quantum interconnects across heterogeneous platforms, essential for enabling scalable modular and cross-platform quantum systems.

While achieving high-fidelity, high-bandwidth transduction still requires research and engineering efforts, it has recently become a major institutional and experimental focal point. Significant advances in various microwave-optical interfaces~\cite{fan2018superconducting, han2020cavity, mirhosseini2020superconducting, zhong2022microwave, meesala2024quantum, sahu2023entangling, zhang2024entanglement, huang2022two, beukers2024remote, dhordjevic2021entanglement} demonstrate a clear and promising path of progress. Crucially, the recent DARPA HARQ program~\cite{darpa2025harq} has issued a major solicitation explicitly targeting the development of NA-SC quantum interconnects. The program's ambitious benchmarks---$10$MHz transmission rates and $99.9\%$ fidelity---signal a significant research thrust intended to solve this challenge within the next few years. These initiatives highlight the need for strategic HQA designs that can effectively harness this interconnect technology as it becomes viable.

\section{Heterogeneous Quantum ISA (HQ-ISA)}
\label{sec:ISA}

This section defines an instruction set architecture for heterogeneous quantum systems (HQ-ISA). Concrete costs (latency, throughput, resource usage) are instantiated in Sec.~\ref{sec:arch_modeling}, followed by an architectural analysis (Sec.~\ref{sec: arch analysis}). We group operations into four categories---\textbf{Compute}, \textbf{Memory}, \textbf{MSF}, and \textbf{Transport (NA--SC)}, summarized in Table~\ref{tab:hqisa}. 

\noindent\textbf{Compute.} 
The compute instructions distinguish NA and SC: NA implements a gate layer via transversal gates (\texttt{NA\_GBC\_LAYER(spec,r)}), while SC implements a PPM layer via lattice surgery (\texttt{SC\_PBC\_PPM(spec,r)}). 
Both carry an implementation specification \texttt{spec} (used for cost evaluation) and include configurable \texttt{r} SM rounds. 
T-layers consume \texttt{m} magic states from the buffer when available via \texttt{MAGIC\_TAKE(m)}; otherwise a preceding \texttt{FENCE\_MAGIC} waits until they arrive. Concretely, a T-gate layer with \(k_T\) T gates issues \texttt{MAGIC\_TAKE(\(k_T\))}, while a PPM layer on SC that always consumes one magic state issues \texttt{MAGIC\_TAKE(1)}. Typical execution sequences are like:
\begingroup
\small  
\setlength{\jot}{0.25ex} 
\begin{align*}
  & \texttt{FENCE\_MAGIC},\ \texttt{MAGIC\_TAKE}(k_T), \texttt{NA\_GBC\_LAYER}(\texttt{spec},\,1);\\[2pt]
  & \texttt{FENCE\_MAGIC},\ \texttt{MAGIC\_TAKE}(1),\ \texttt{SC\_PBC\_PPM}(\texttt{spec},\,d_{\text{surf}});
\end{align*}
\endgroup
where transversal gate layer requires only 1 SM round and lattice-surgery-based PPM layer requires $d_{\text{surf}}$ SM rounds.

\noindent\textbf{Memory.}
At each layer, only the logical qubits that participate in operations need to be in the compute region; inactive qubits can remain in memory to reduce the overall space footprint. Because the active set evolves across layers, the system performs an \emph{interchange}: newly active qubits are loaded while no-longer-active qubits are stored. This is expressed by \texttt{LDPC\_LOAD(qset)} and \texttt{LDPC\_STORE(qset)}, which move logical qubits between compute and memory. 
\texttt{BUF\_RESERVE($Q$)} sizes the logical swap buffer at the memory–compute boundary on NA-side that facilitates this process. The size of this buffer affects how long it takes to finish the logical swap.
If memory and compute reside on different modalities, cross-modality transfer of logical qubits is made explicit via \texttt{TRANS\_LQ(dir, qset)} (see Transport). 

\noindent\textbf{MSF.} MSF may locate on multiple hosts, continuously producing magic states and sending them to the compute for consumption. The instruction \texttt{MSF\_START(host,k)} activates \(k\) parallel MSF on the designated host, where completion is stochastic due to its repeat-until-success procedure. The availability on the compute side is reflected through \texttt{MAGIC\_TAKE} after possible delivery over the interconnect, if MSF and compute are located on different modalities.

\noindent\textbf{Transport (NA--SC).}
In HQA, cross-modality traffic arises either between memory and compute (MCSep) or between MSF and compute (MAcc). 
We model both cases explicitly: 
\texttt{SEND\_MAGIC(host,m)} transmits \(m\) magic states from a specific MSF \texttt{host} to the compute-side buffer (used only when MSF and compute are on different modalities), and \texttt{RECV\_MAGIC(host,m)} acknowledges arrival and makes those tokens visible in the buffer. 
\texttt{TRANS\_LQ(dir,qset)} transfers a logical-qubit set across modalities in direction \(\texttt{dir} \in \{\text{NA--SC}, \text{SC--NA}\}\) and completes when the qubits are available at the destination.

\section{Cost Modeling}
\label{sec:arch_modeling}

Following the HQ-ISA (Sec.~\ref{sec:ISA}), we model costs for four categories, adaptable to different HQA designs and FTQC configurations. 
Each category has a parameterized cost model (Sec.~\ref{subsec:arch_compute}–\ref{subsec:arch_msf}), after which we compose them to obtain end-to-end runtime, qubit footprint, and breakdowns (Sec.~\ref{subsec:arch_compose}). 
Default parameters are provided for concreteness but remain reconfigurable; homogeneous designs arise as special cases. 

\subsection{Compute Modeling}
\label{subsec:arch_compute}

We adopt surface codes for compute (both NA and SC) due to their well-established logical operations. We models two execution schemes--GBC on NA and PBC on SC, specifying the execution time and qubit resource required. This framework also extends to other schemes and qubit modalities.

\begin{figure}[!ht]
    \centering
    \vspace{-6pt}
    \includegraphics[width=0.48\textwidth]{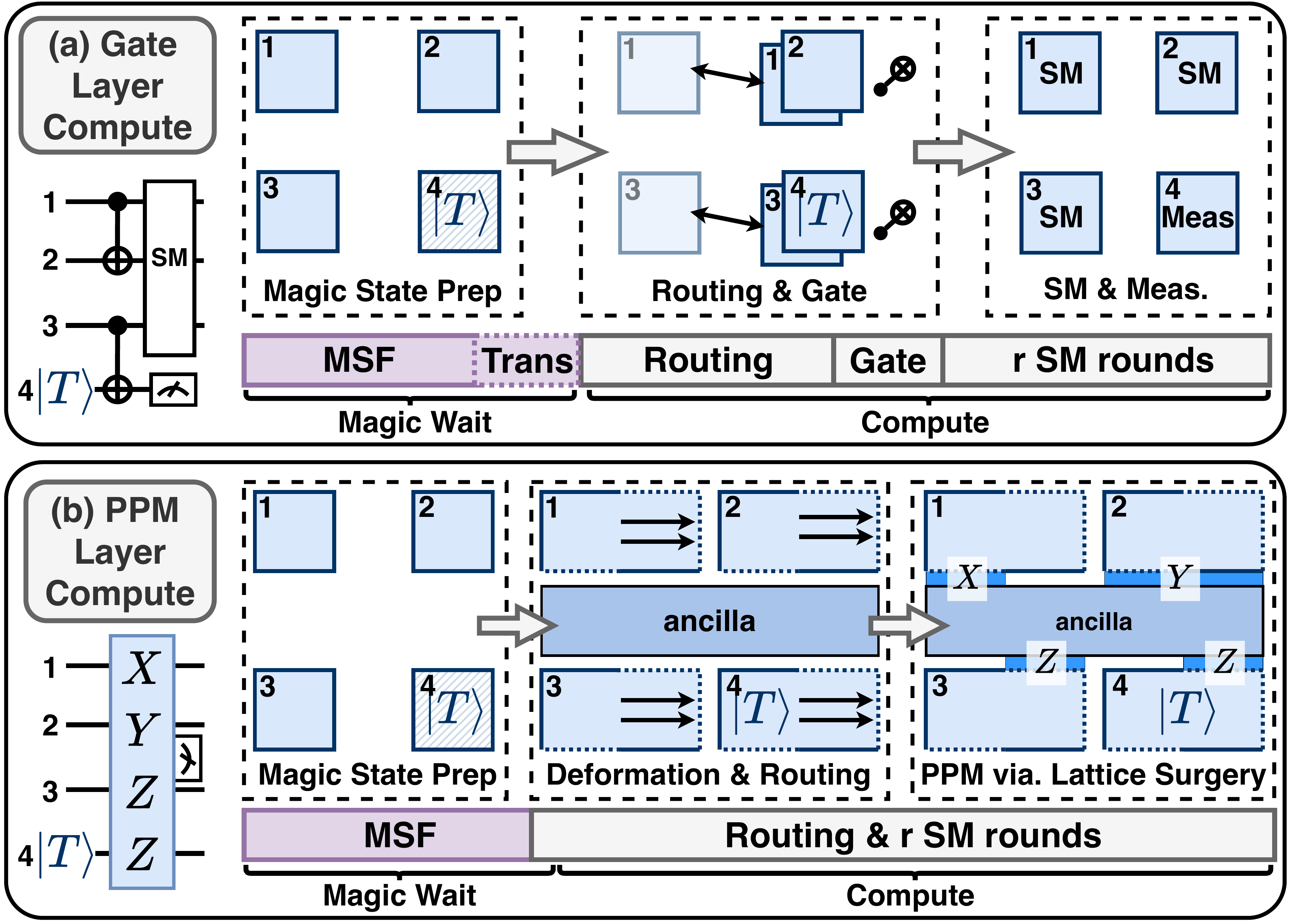}
    \vspace{-6pt}
    \caption{Compute modeling for GBC and PBC schemes.}
    \label{fig:compute modeling} 
    \vspace{-6pt}
\end{figure}

\begin{figure*}[!ht]
    \centering
    \includegraphics[width=.99\textwidth]{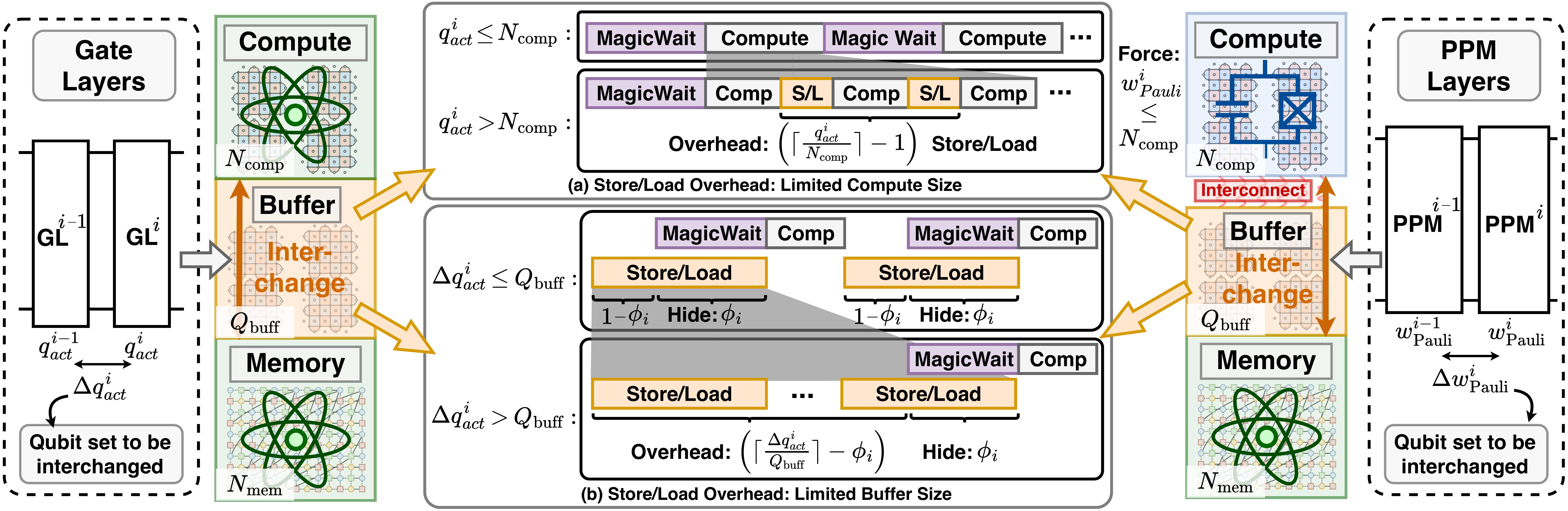}
    \vspace*{-6pt}
    \caption{Modeling of store/load overhead due to limited compute and buffer size.}
    \vspace*{-10pt}
    \label{fig: SL model}
\end{figure*}

\noindent\textbf{\texttt{NA\_GBC\_LAYER(spec,r).}} \textit{Execution Time.} Fig.~\ref{fig:compute modeling}(a) gives the typical execution pipeline of a gate layer with T gates on NA. The total execution time $t_{\text{GL}}(\texttt{spec},r)$ consist of several major components:
\begin{align} \label{GL time 1}
    \begin{split}
        t_{\text{GL}}(\texttt{spec}, r) \ &= \ W_{\text{Magic}}\cdot \mathbf{1}_{\text{T-layer}}+t_{\text{route}}^{\text{NA}}(\texttt{spec})\\
        & +\ t_{\text{gate}}^{\text{NA}}(\texttt{spec}) \ + \ r\cdot t_{\text{SM}}^{\text{NA}}, 
    \end{split}
\end{align}
where $W_{\text{Magic}}$ is the wait for magic-state availability, and $\mathbf{1}_{\text{T-layer}}{=}1$ only if the layer contains $T$ gates (otherwise $0$). The term $t_{\text{route}}^{\text{NA}}(\texttt{spec})$ accounts for physically co-locating interacting logical qubits (e.g., via tweezer moves for two-qubit gates)~\cite{bluvstein2024logical}; $t_{\text{gate}}^{\text{NA}}(\texttt{spec})$ is the transversal gate time; and $r\cdot t_{\text{SM}}^{\text{NA}}$ is the time for $r$ SM rounds. We propose a realistic execution model aligned with state-of-the-art NA estimates~\cite{ruan2025powermove, bluvstein2024logical, zhou2025resource}:

\noindent\textit{(1) Logical Routing.} We assume logical qubits (distance-$d$ surface code patches) are placed on a $\lceil n \rceil \times \lceil n \rceil$ grid for a $n$-logical-qubit benchmark, with $10\,\mu m$ physical qubits spacing. This yields a overall layout spanning $10d\lceil n\rceil \times 10d\lceil n\rceil\,\mu m^2$ with worst-case movement distance $\sqrt{2}\cdot10d \lceil n \rceil\ \mu m$. We estimate per-layer routing time $t_{\text{route}}^{NA}$ by the worst-case move time $t_{\text{Move}}$, using the formula~\cite{ruan2025powermove, tan2025compilation}
\[
t_\text{Move} = \left( \frac{2 \cdot 10d \lceil n \rceil \cdot \sqrt{2}}{2750} \right)^{1/2}\, ms.
\]
For instance, with $d=15$ and $n=100$, $t_\text{Move} \approx 1.24ms$. We assume sufficient tweezer parallelism so all required patch motions for a layer complete within a single $t_{\text{Move}}$.

\noindent\textit{(2) SM rounds.} One SM round time comprises four movements for enabling CNOTs (roughly $400\,\mu s$~\cite{fowler2012surface, gidney2021stim}), two steps of physical $H$ gates ($1\mu s$~\cite{bluvstein2024logical}) and a measurement stage ($500\mu s$~\cite{bluvstein2024logical}), in total around $900\mu s$. With transversal execution on NA, a single round suffices for FT protection \cite{zhou2025low, zhou2025resource}, so we use $r${=}1 unless stated otherwise.

\noindent\textit{(3) Gate Time.} The transversal gates consist of physical gates that are implemented in parallel and are typically fast $<1\,\mu s$~\cite{bluvstein2024logical}, negligible compared to routing and SM rounds.
Overall, a typical GBC layer incurs $\sim$1–3 $ms$ of latency, plus magic state supply delay $W_{\text{Magic}}$ if needed.

\noindent\textit{Space Overhead.} Suppose there are $N$ logical qubits encoded in distance-$d$ surface codes in the compute region. The total qubit resource is then $2Nd^2$, with each of the $N$ logical qubits occupying $2d^2$ physical qubits.

\noindent\textbf{\texttt{SC\_PBC\_PPM(spec,r)}}. \textit{Execution Time.} Fig.~\ref{fig:compute modeling}(b) shows the typical execution pipeline of a PPM layer on SC. The total execution time $t_{PPM}(\texttt{spec},r)$ is:
\begin{align}{\label{PPM time}}
    \begin{split}
        t_{\text{PPM}}(\texttt{spec}, r)=W_{\text{Magic}}+r\cdot t_{\text{SM}}^{\text{SC}}, 
    \end{split}
\end{align}
Notably, the code deformation and routing path preparation stage only involves initialization of physical qubits so it does incur time overhead~\cite{litinski2019game}, and the later syndrome measurement stage requires $d_{\text{surf}}$ rounds to guarantee fault-tolerance. Therefore, we assume $r=d_{\text{surf}}$ unless otherwise stated. Each SM round on SC typically takes around $1\,\mu s$~\cite{beverland2022assessing, gidney2021factor, acharya2024quantumerrorcorrectionsurface}.

\noindent\textbf{\texttt{MAGIC\_TAKE(m)}/\texttt{FENCE\_MAGIC}.} They are used for adjusting the execution pace according to magic states availability and do not induce overhead.

The modeling of this subsection assumes the compute region is sufficiently large to hold all active qubits within a layer. Next subsection will discuss the extra time overhead when the compute size is limited.

\subsection{Memory–Compute Interaction Modeling}
\label{subsec:arch_mcsep}

In the memory–compute separation design, we use qLDPC codes for memory and surface codes for compute. 
We first specify qubit \emph{resource allocation} for memory, compute, and the logical buffer, then model the time overhead under constraints on the size of compute and buffer.

\noindent\textbf{Resource allocation.}
\textit{Memory (qLDPC).}
We instantiate memory blocks using the $[[288,12,18]]$ qLDPC code~\cite{bravyi2024high}, leveraging NA’s flexible connectivity~\cite{bluvstein2024logical,bluvstein2025architectural}. 
Each block stores $12$ logical qubits, occupies $576$ physical qubits (including ancilla), has code distance $d_{\text{qLDPC}}$, and achieves logical error rate $\sim 10^{-9}$~\cite{yoder2025tour}. 
A logical processing unit (LPU) of $158$ ancilla per block enables PPM-based store/load with $d_{\text{qLDPC}}$ SM rounds~\cite{yoder2025tour,stein2025hetec,he2025extractors}. $B$ memory blocks give capacity of $N_{\text{mem}}{=}12B$ logical qubits and requires $n_{\text{phys}}^{\text{mem}}{=}(576+158)B$ physical qubit in total.

\noindent\textit{Compute \& Logical Buffer (surface code).} As explained in Sec.~\ref{subsec:arch_compute}, a compute region (resp. logical buffer) with $N_{\text{comp}}$ (resp. $Q$) logical qubits requires $2d_{\text{surf}}^2 N_{\text{comp}}$ and $8d_{\text{surf}}^2 N_{\text{comp}}$ (resp. $2d_{\text{surf}}^2 Q$ and $8d_{\text{surf}}^2 Q$) qubit resources for NA and SC. The limited size of compute or buffer will incur store/load overhead, requiring instructions \texttt{LDPC\_LOAD(qset)}, \texttt{LDPC\_STORE(qset)} to interchange between memory and compute based on the evolution of the active qubit set in each layer, as explained below.

\noindent\textbf{Time Overhead: Limited Compute Size.} For GBC scheme, a limit compute size will force a gate layer to be implemented in multiple steps. As Fig.\ref{fig: SL model} (left) shows, assuming the $i$-th gate layer have $q_{\text{act}}^i$ active qubits and the compute size is $N_{\text{comp}}$, then it takes $s_i=\lceil q_{\text{act}}^i/N_{\text{comp}}\rceil$ store/load processes to interchange all the logical qubits needed for performing all the gates in the layer. This will require extra $(s_i-1)$ times of transversal gates + SM rounds ($(1+r)$ cycles), and $(s_i-1)$ extra batches of store/load time overhead (Fig.\ref{fig: SL model}(a)), each taking $d_{\text{qLDPC}}$ cycles. For PBC scheme (Fig.~\ref{fig: SL model}, right), compute size must be no smaller than the PPM weight: $N_{\text{compute}} \geq \max_{1\leq i\leq D}(w_{\text{Pauli}}^i)$, where $w_{\text{Pauli}}^i$ is the Pauli weight of the $i$-th PPM layer (i.e., the number of active qubits); otherwise, the corresponding PPMs cannot be performed. This constraint can significantly increase SC compute overhead for applications with high Pauli weights.

\noindent\textbf{Time Overhead: Limited Buffer Size.} Let $\Delta q_{\text{act}}^i$ denote the number of logical qubits to be swapped for the $(i-1)$-th to $i$-th layer transition (including both the storing-in and loading-out qubits) and $Q_{\text{buff}}$ the reserved buffer capacity. When $\Delta q_{\text{act}}^i \leq Q_{\text{buff}}$, we assume the latency can be proportionally hidden by pipelining, specified by the parameter $\phi_i$; if $\Delta q_{\text{act}}^i>Q_{\text{buff}}$, it will incur $(\lceil \Delta q_{\text{act}}^i/Q_{\text{buff}}\rceil-\phi_i)$ additional store/load processes (see Fig.~\ref{fig: SL model}(b)), each taking $d_{\text{qLDPC}}$ cycles. 

Combining these two sources (Fig.\ref{fig: SL model}), the total time overhead (counted in cycles) becomes:
\begin{align}{\label{C overhead}}
\begin{split}
    C_{\text{over}} = \ & \mathbf{1}_{\{q_{\text{act}}^i>N_{\text{comp}}\}} (s_i-1)\left[(1+r)+\lceil \tfrac{N_{\text{comp}}}{Q_{\text{buff}}}\rceil d_{\text{qLDPC}}\right]\\
    + \ & \mathbf{1}_{\{\Delta q_{\text{act}}^i >Q_{\text{buff}}\}}  \left(\lceil \tfrac{\Delta q_{\text{act}}^i}{Q_{\text{buff}}}\rceil - \phi_{\text{hide}}\right)\cdot d_{\text{qLDPC}} \\
    + \ & \mathbf{1}_{\{\Delta q_{\text{act}}^i <Q_{\text{buff}}\}}  \left(1 - \phi_{\text{hide}}\right)\cdot d_{\text{qLDPC}} 
\end{split}
\end{align}
The above formula targets for gate layers (GBC). For PBC, a PPM layer is infeasible unless $N_{\text{comp}}\!\ge\! w_{\text{Pauli}}^i$; we therefore set $N_{\text{comp}}=\max_{1\le i\le D} w_{\text{Pauli}}^i$, which eliminates the first term in (\ref{C overhead}). Replacing the $q_{\text{act}}^i$ in the rest of terms by $w_{\text{Pauli}}^i$ will give the time overhead for PBC.

\subsection{Magic-state factory and NA--SC interconnect}
\label{subsec:arch_msf}

\noindent\textbf{Factory throughput and queueing.}
Each MSF is modeled as a stochastic throughput source. 
With per-attempt cycles $C_{\text{att}}$, success probability $p$, and hardware cycle time $t_{\text{cycle}}^{\text{mod}}$, the time per magic state and the aggregate throughput are
\[
t_{\text{MS}}=C_{\text{att}}\cdot t_{\text{cycle}}^{\text{mod}}/p,\qquad 
\mu_{\text{MSF}}=k/t_{\text{MS}},
\]
for $k$ parallel factories. 
By default we instantiate magic-state cultivation (MSC)~\cite{gidney2024magic} due to its practically compact size: $463$ physical qubits, $24$ cycles per attempt, and $1\%$ success per attempt to reach target infidelity $10^{-9}$–$10^{-10}$, yielding $\approx\!2400$ cycles per state (wall time $t_{\text{MS}}$ follows the hardware cycle, e.g., $t^{\mathrm{SC}}_{\mathrm{cycle}}{=}1\,\mu\mathrm{s}$; $t^{\mathrm{NA}}_{\mathrm{cycle}}{=}1\,\mathrm{ms}$). 
Consumption follows ISA blocking semantics: \texttt{MAGIC\_TAKE($m$)} blocks until $m$ magic states have \emph{arrived} in the compute-side buffer. We provision buffer capacity of $Q_M$ logical qubits, requiring $2Q_Md_{\text{surf}}^2$ physical qubits, to store magic states for consumption.

\noindent\textbf{Magic-state delivery over the NA--SC link.}
When MSF and compute lie on different modalities, magic states transfer through the interconnect and then become visible in compute (\texttt{SEND\_MAGIC}/\texttt{RECV\_MAGIC}). 
We model the path as two pipelines: the factory produces at rate $\mu_{\text{MSF}}$ and the link delivers at rate $\mu_{\text{link}}$. The effective steady delivery is
$\mu_{\text{deliv}}=\min(\mu_{\text{MSF}},\,\mu_{\text{link}})$,
with a one-way latency $t_{\mathrm{MST}}$ that acts as an arrival offset. For a burst of m states,
$$t_{\text{link}}(m)=t_{\mathrm{MST}}+\frac{m}{\mu_{\text{deliv}}}.$$
Loss is accounted for in the effective rate $\mu_{\text{link}}$, reducing the throughput. Correctness is unchanged since magic states do not encode program information.

\noindent\textbf{Logical-qubit transport.}
In MCSep, we assume program states move via fault-tolerant teleportation over \emph{verified} (heralded) Bell pairs~\cite{ramette2024fault, sinclair2025fault}. With heralding, failures are primarily caught during Bell-pair generation and retried, so correctness is preserved to first order and the impact appears as added delay rather than silent errors. For evaluation we adopt a minimal, optimistic transfer model: each invocation transfers up to $B$ logical qubits and costs a fixed $t_{\mathrm{LQT}}$ (``Logical Qubit Transport''). 
Thus moving $n$ qubits completes in
\[
t_{\text{batch}}^{\text{LQ}}(n)=\Big\lceil \frac{n}{B} \Big\rceil\, t_{\mathrm{LQT}}.
\]
Protocol details (pair generation, purification, acknowledgements, residual loss) are absorbed into $t_{\mathrm{LQT}}$ and $B$. In particular, the homogeneous version of MCSep (NA memory + NA compute) requires no cross-modality link, thus avoids link-reliability assumptions while still delivering space savings via qLDPC memory, giving a more practical near-term scheme.

\subsection{End-to-end composition}
\label{subsec:arch_compose}
Given a layered FT workload $\{1,\ldots,D\}$ and a concrete HQA configuration, we estimate resources by composing the per-category models.

\noindent\textbf{Execution Time.}
For each layer $i$, the runtime is the \emph{base compute time} from Sec.~\ref{subsec:arch_compute} (Formula (\ref{GL time 1}, \ref{PPM time})), adjusted by overhead induced by MCSep (store/load, limited compute size) if applicable (Formula (\ref{C overhead})). The magic state wait time $W_{\text{Magic}}$ for each layer is specified in Sec.~\ref{subsec:arch_msf} by adding the MSF time $t_{\text{MS}}$ and transport time $t_{\text{link}}$. If memory and compute locate on different modalities, the logical qubit transport time $t_{\text{batch}}^{\text{LQ}}(n)$ needs to be added. 

\noindent\textbf{Qubit Resource.}
Physical qubits are tallied additively as
\[
n_{\text{phys}} \;=\; n_{\text{compute}} \;+\; n_{\text{memory}} \;+\; n_{\text{MSF}} \;+\; n_{\text{buff}},
\]
with $n_{\text{compute}}$ from the compute, $n_{\text{memory}}$ from the qLDPC memory block if applicable, $n_{\text{MSF}}$ from the number of parallel factories, and $n_{\text{buff}}$ from the configured swap/magic buffers. 
The resource estimation can also report a breakdown across compute, magic wait, store/load, and transport.
\section{Architectural Analysis}
\label{sec: arch analysis}

This section turns the HQ-ISA cost model into design insight via tractable, proof-of-principle analyses. We first analyze the \emph{speedup of MAcc} (Sec.~\ref{subsec:MagicAcc analysis}), isolating Clifford time from magic-state stalls and identifying the key drivers across benchmark features, FTQC protocols, and hardware capabilities. We then quantify the \emph{space–time trade-off} of \emph{MCSep} (Sec.~\ref{subsec:MCSep analysis}) by coupling qLDPC packing, compute-region size, swap-buffer capacity, and logical-transfer cost into the memory-interaction term. The resulting expressions expose phase boundaries and rules of thumb that map workload properties to the preferred heterogeneous design. Combined designs and broader trade-offs are examined in Sec.~\ref{sec: evaluation}.

\subsection{MAcc Analysis}
\label{subsec:MagicAcc analysis}
\begin{figure}[!ht]
    \centering
    \vspace{-6pt}
    \includegraphics[width=0.48\textwidth]{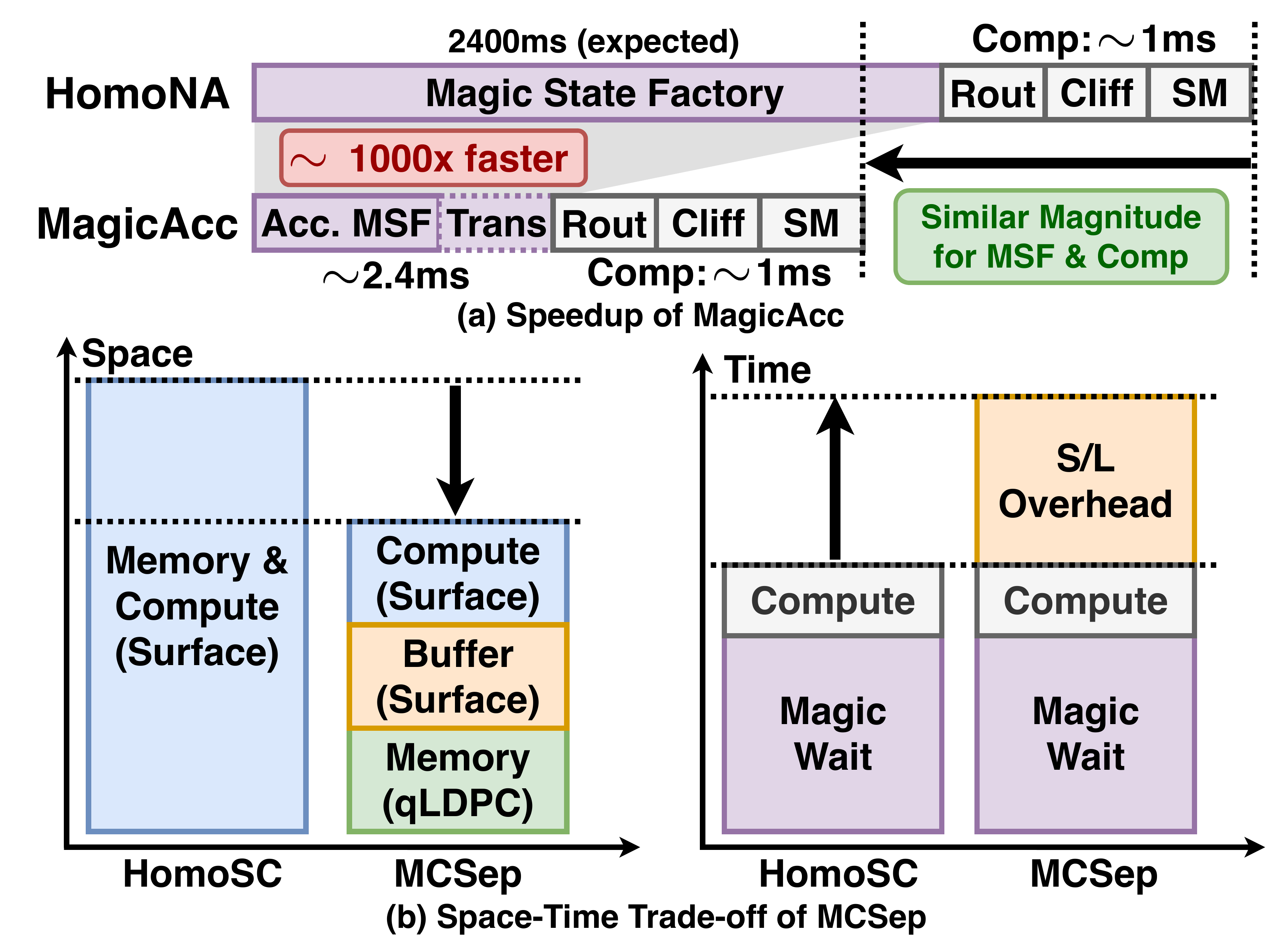}
    \vspace{-10pt}
    \caption{Architecture analysis for (a) MAcc speedup and (b) MCSep space-time trade-off.}
    \label{fig:arch analysis} 
    \vspace{-10pt}
\end{figure}

We analyze a canonical \emph{MAcc} configuration—MSF on SC and compute on NA (all surface code)—against a homogeneous NA baseline. Time is counted in \emph{cycles} for both compute and MSF: a NA gate layer without $T$ uses $(1{+}r)$ NA cycles (one routing plus $r$ SM rounds), and sub-$\mu$s gate times are neglected. A $T$-consuming layer adds the MSF latency of $C_{\text{MSF}}$ cycles (measured on the host modality), and for MAcc includes a cross-modal delivery penalty $t_{\mathrm{MST}}$. While exact per-cycle times can differ across operations, they are of the same order within a modality; treating them uniformly yields a fair first-order estimate, as adopted in prior work~\cite{litinski2019game, gidney2021factor, beverland2022assessing}. Wall time then follows by multiplying cycles by the hardware cycle time (e.g., $T_{\text{cycle}}^{\text{NA}}$, $T_{\text{cycle}}^{\text{SC}}$). The intuition of why the MAcc leads to system-level gain is illustrated in Fig.\ref{fig:arch analysis}(a), where the MSF on slow NA devices are accelerated by the $\sim 1000$x faster SC device. The quantitative analysis of overall speedup is presented below.

\noindent\textbf{Amdahl-style speedup.}
Let $T_{\text{cycle}}^{\text{NA}}$ and $T_{\text{cycle}}^{\text{SC}}$ be cycle times, $C_{\text{MSF}}$ the expected MSF cycles per state, $D$ the number of layers, and $N_T$ the number of $T$-layers. Then
\[
\begin{aligned}
t_{\text{C-layer}}=(1{+}r) & T_{\text{cycle}}^{\text{NA}},  \ t_{\text{T-layer}}^{\text{HomoNA}}=t_{\text{C-layer}}+C_{\text{MSF}}T_{\text{cycle}}^{\text{NA}},\\
t_{\text{T-layer}}^{\text{MAcc}}&=t_{\text{C-layer}}+C_{\text{MSF}}T_{\text{cycle}}^{\text{SC}}+t_{\mathrm{MST}}.
\end{aligned}
\]
The total-time ratio (speedup) is
\[
\mathrm{Speedup}
=\frac{(D{-}N_T)\cdot t_{\text{C-layer}}+N_T\cdot t_{\text{T-layer}}^{\text{HomoNA}}}
{(D{-}N_T)\cdot t_{\text{C-layer}}+N_T\cdot t_{\text{T-layer}}^{\text{MAcc}}}.
\]
Define the hardware speed ratio $S=T_{\text{cycle}}^{\text{NA}}/T_{\text{cycle}}^{\text{SC}}$, T-layer ratio $r_T{=}N_T/D$, the Clifford–MSF gap $\rho_{\text{MS}}=\tfrac{C_{\text{MSF}}}{1{+}r}$, and the normalized transport penalty $P_{\text{Trans}}=\tfrac{t_{\mathrm{MST}}}{C_{\text{MSF}}T_{\text{cycle}}^{\text{SC}}}$. After algebra,
\begin{equation}
    \mathrm{Speedup}
=1+\frac{S-1-P_{\text{Trans}}}
{\,1+\tfrac{1}{r_T}\,\tfrac{1}{\rho_{\text{MS}}}\,S+P_{\text{Trans}}\,}
\end{equation}
which isolates four drivers:
(i) larger hardware speedup ($S$) helps;
(ii) larger $T$-layer ratio ($r_T$) benefits;
(iii) larger MSF\,/\,Clifford cycle gap ($\rho_{\text{MS}}$) helps; while
(iv) higher delivery penalty ($P_{\text{Trans}}$) erodes gains.

\noindent\textbf{Insights.} This mirrors Amdahl’s law. The gain is not set by hardware speed alone ($S$), but also by how much of the program can benefit and by how severe that part is: the fraction of $T$-layers ($r_T$) and the Clifford–vs.–magic gap ($\rho_{\text{MS}}$). Higher $r_T$ means more opportunities to accelerate; larger $\rho_{\text{MS}}$ means the bottlenecked portion dominates, so speeding it up yields significant system-level gains.

\noindent\textbf{Upper bound.}
We consider the favorable limit: every layer consumes magic states ($r_T\!\to\!1$), transport is free ($P_{\text{Trans}}\!=\!0$), and SC is arbitrarily fast ($S\!\to\!\infty$). From the closed form,
\[
\mathrm{Speedup}_{\max} \;=\; \lim_{S\to\infty}\!\Bigl[\,1+\frac{S}{1+\tfrac{1}{\rho_{\text{MS}}}S}\Bigr]
\;=\; 1+\rho_{\text{MS}}.
\]
\emph{Interpretation.} The ceiling is set by the \emph{Clifford–vs.–magic gap} $\rho_{\text{MS}}$, not by raw hardware speed $S$ or the $T$-layer fraction $r_T$. Even with “infinitely fast” SC and no transport cost, speedup saturates at $1{+}\rho_{\text{MS}}$; if $\rho_{\text{MS}}$ is small, the payoff is intrinsically limited, whereas a large $\rho_{\text{MS}}$ provides substantial headroom.

\noindent\textbf{Design implications.} The above analysis shows that \emph{MAcc} is most attractive when $\rho_{\text{MS}}$ is large—e.g., cultivation protocol ($C_{\text{MSF}}{=}2400$ cycles/state) paired with single-round transversal Clifford layers ($r{=}1$), yielding $\rho_{\text{MS}}{=}\tfrac{C_{\text{MSF}}}{1{+}r}{=}1200$. Pairing modalities with a large speed ratio $S$ (e.g., NA–SC or TI–SC, $\sim10^3$) or targeting workloads with higher $r_T$ moves performance closer to the upper bound. Finally, as long as link latency is on the order of MSF time or smaller ($P_{\text{Trans}}\!\lesssim\!1$), transport does not become the system bottleneck and MAcc retains its advantage.

\subsection{MCSep Analysis}
\label{subsec:MCSep analysis}
We study how the sizing of the compute ($N_{\text{comp}}$), qLDPC memory ($N_{\text{mem}}$), and the logical swap buffer ($Q_{\text{buffer}}$) drives the space–time trade-off in \emph{MCSep} (Fig.\ref{fig:arch analysis}(b)). The time overhead due to limited size of compute and logical buffer is given in equation (\ref{C overhead}). We focus on the gate-layer case because the PPM layer case can be similarly discussed.

We assume the layered workload have active-qubit counts $\{q_{\text{act}}^i\}_{i=1}^D$ and inter-layer changes $\{\Delta q_{\text{act}}^i\}_{i=2}^D$ (the number of qubits entering/leaving compute).
Suppose $N_{\text{comp}}$ and $Q_{\text{buff}}$ fall on the $\alpha$- and $\beta$-quantiles of the empirical distribution of $\{q_{\text{act}}^i\}_{i=1}^D$ and $\{\Delta q_{\text{act}}^i\}_{i=2}^D$:
\[
\Pr(q_{\text{act}}^i \le N_{\text{comp}})=\alpha,\qquad
\Pr(\Delta q_{\text{act}}^i \le Q_{\text{buffer}})=\beta,
\]
Taking expectations of the time overhead (equation (\ref{C overhead})), lower-bounding the tail event ($s_i\geq 2$ and $\lceil \tfrac{\Delta q_{\text{act}}^i}{Q_{\text{buffer}}}\rceil\geq 2$ when the ahead indicator function holds), and some algebra give
\begin{align}
\begin{split}
    & \qquad\mathbb{E}[C_{\text{over}}] \geq (1-\alpha)\cdot(1+r) \ + \ \\
    & \left[(1-\alpha)\cdot \lceil \tfrac{N_{\text{comp}}}{Q_{\text{buffer}}}\rceil + (2-\beta-\phi_{\text{hide}})\right]\cdot d_{\text{qLDPC}}
\end{split}
\end{align}
The results shows that larger $\alpha$, $\beta$ shrink the tails that trigger memory-compute interchange, directly lowering stalls.

\noindent\textit{Space–time trade-off.} We note that when $N_{\text{comp}}\ge \max_i \{q_{\text{act}}^i\}$, $Q_{\text{buffer}}\ge\max_i \{\Delta q_{\text{act}}^i\}$ (which implies $\alpha$=$\beta$=1), and $\phi_{\text{hide}}\!=\!1$, it saturates the lower bound of $0$, meaning zero time overhead. However, this requires large compute and buffer region and potentially offsets qLDPC’s space advantage. Conversely, aggressive space caps ($K = \frac{N_{\text{total}}}{N_{\text{comp}}}\!\gg\!1$) gives a small compute region, reducing footprint but increasing the time penalties. Relative to an ideal per-layer cost of $(1{+}r)\;+\;r_T\cdot (t_{\mathrm{MSF}}+t_{\mathrm{MST}})/{T_{\mathrm{cycle}}^{\mathrm{NA}}}$, the memory-interaction overhead becomes significant when $(t_{\mathrm{MSF}}+t_{\mathrm{MST}})/{T_{\mathrm{cycle}}^{\mathrm{NA}}}$ is small—precisely the case in MAcc (fast SC MSF plus modest transport). Thus, in MCSep+MAcc, store/load can become a first-order contributor to runtime unless $N_{\mathrm{comp}}$ and $Q_{\mathrm{buff}}$ are provisioned to limit store/load.

\noindent\textbf{Design Implications.} MCSep is most attractive when (i) with NA compute, the active set per layer and its inter-layer churn are small (low $\{q_{\text{act}}^i\}$, low $\{\Delta q_{\text{act}}^i\}$); or (ii) with SC compute, PPM layers have low Pauli weights (low $\{w_{\text{Pauli}}^i\}$). Otherwise, keeping store/load overhead low demands large $N_{\text{comp}}$ and $Q_{\text{buffer}}$, which can erode the qLDPC memory’s space advantage. For scheme choice: workloads with high active-qubit counts but modest Pauli weights tend to favor SC–PBC, whereas workloads with sparse activity or high Pauli weights favor NA–GBC.



\section{Evaluation} \label{sec: evaluation}

\subsection{Experiment Setup}\label{subsec: exp setup}

\noindent\textbf{Baseline Designs.} We compare against a suite of baselines reflecting realistic and forward-looking implementation choices. They are first classified into two groups based on their QEC code usage (whether MCSep is adopted). \textbf{Group 1: Surface Code Only (No MCSep).} All functional modules, including memory, compute, and MSF, use surface codes. Despite their substantial space overhead, surface codes remain the most mature and scalable QEC approach for reaching arbitrarily low LERs~\cite{litinski2019game, litinski2019magic, gidney2021factor, beverland2022assessing}. \textbf{Group 2: qLDPC + Surface Code (MCSep).} Here, qLDPC codes are used exclusively for memory, while other modules continue to rely on surface codes. This reflects the original use case of qLDPC codes as space-efficient storage~\cite{bravyi2024high, stein2025hetec}.

Each group further explores three hardware configurations: (1) \textit{Homo SC}, (2) \textit{Homo NA}, and (3) \textit{Hete NA–SC}. This gives in total $6$ baseline configurations, summarized in Table~\ref{tab:baseline-summary}. Notably, in Group 1, we include Homo SC (\textbf{SC-SF}) despite its prohibitive space cost that goes beyond SC's scalability, using it as an idealized lower bound on execution time. The Hete NA-SC gives the MAcc design (\textbf{HT-SF-MAcc}), where MSF is overloaded to SC and the rest of the system operates on NA. In Group 2, we exclude Homo SC because qLDPC codes remain difficult to implement on SC platforms due to hardware constraints. However, we consider two heterogeneous configurations: \textbf{HT-MCSep}, where NA supports qLDPC memory and SC supports both compute and MSF (so there's no ``magic state acceleration'' compared to the compute); \textbf{HT-MCSep-MAcc}, where NA supports qLDPC memory and compute, and SC supports MSF (so there's ``acceleration''). Together, these baselines span a broad architectural space, enabling a systematic analysis of space–time trade-offs under both practical and forward-looking assumptions. The execution models used for each configuration have been detailed in Sec.~\ref{sec:arch_modeling}. The homogeneous baseline NA-MCSep with NA qLDPC memory and NA compute is introduced for comparing the speedup from MAcc in MCSep systems.

\vspace{2pt}
\noindent\textbf{Metrics.} We evaluate each baseline using the following system-level metrics:

\noindent\textit{1. Physical qubit count.} Total number of physical qubits used, including memory, compute, MSF, and ancillary components (buffer zones) supporting the system.

\noindent\textit{2. Execution time.} End-to-end runtime measured in seconds for direct comparison, computed following the execution model detailed in Sec.~\ref{sec:arch_modeling}.

\noindent\textbf{Setup and assumptions.} Logical fidelity is not a point of comparison, as we select surface code distance that achieve sufficiently low LER to meet a standard $90\%$ overall success probability across all benchmarks. In the overall resource estimation (Sec.\ref{subsec: overall performance}), we assume one MSF instance and an NA–SC delivery latency $t_{\mathrm{MST}}=10^{-7}\,\mathrm{s}$. The memory–compute interface is workload-tuned: we set the buffer size $Q_{\text{buff}}$ to the 0.95-quantile of $\{\Delta q_{\text{act}}^i\}$ for GBC and to the 0.8-quantile of $\{\Delta w_{\text{act}}^i\}$ for PBC. Compute sizes follow scheme-specific constraints: for PBC, $N_{\text{comp}}=\max_i w_{\text{Pauli}}^i$ so every PPM is feasible; for GBC, $N_{\text{comp}}=\min\!\bigl(Q_{0.5}(\{q_{\text{act}}^i\}_{i=1}^D),\, N_{\text{total}}/3\bigr)$ to preserve space savings without excessive swap overhead (Sec.\ref{subsec:MCSep analysis}), where $Q_{\alpha}(\cdot)$ means the $\alpha$-quantile. We use these settings for headline results and sweep them in the design-space analysis (Sec.~\ref{subsec: ablation}).

\begin{table}[htbp]
\vspace{-6pt}
\centering
\caption{Summary of baseline architectures. Proposed NA–SC heterogeneous architectures are shown in bold.}
\label{tab:baseline-summary}
\small
\setlength{\tabcolsep}{3.5pt}
\renewcommand{\arraystretch}{1.25}
\vspace{-6pt}
\begin{tabular}{|c||c|c|c|}
\hline
\textbf{Group} & \textbf{Scheme Name} & \textbf{Architecture Summary} \\
\hline
\multirow{3}{*}{\makecell[c]{\\ \textbf{1. Surface} \\ \textbf{Code Only} \\ \textbf{(No MCSep)}}} 
  & \makecell[c]{NA-SF~\cite{zhou2025resource}}  
  & \makecell[l]{NA surface code for memory, \\ compute, and MSF} \\
\cline{2-3}
  & \makecell[c]{SC-SF~\cite{litinski2019game, gidney2021factor}} 
  & \makecell[l]{SC surface code for memory,\\  compute, and MSF} \\
\cline{2-3}
  & \makecell[c]{\textbf{HT-SF-MAcc}}  
  & \makecell[l]{NA compute and SC MSF, \\ both with surface code} \\
\hline
\multirow{3}{*}{\makecell[c]{\\ \textbf{2. qLDPC+} \\ \textbf{Surface} \\ \textbf{(MCSep)}}} 
  & \makecell[c]{NA-MCSep~\cite{xu2024constant}}  
  & \makecell[l]{NA qLDPC memory, NA \\ surface code compute + MSF} \\
\cline{2-3}
  & \makecell[c]{\textbf{HT-MCSep}}  
  & \makecell[l]{NA qLDPC memory, \\ SC compute; SC MSF} \\
\cline{2-3}
  & \makecell[c]{\textbf{HT-MCSep}  \\ \textbf{-MAcc}}  
  & \makecell[l]{NA qLDPC memory, \\ NA compute; SC MSF} \\
\hline
\end{tabular}
\vspace{-6pt}
\end{table}

\vspace{2pt}
\noindent\textbf{Benchmark Selection.} We select a representative suite of benchmarks drawn from~\cite{li2023qasmbench, quetschlich2022mqt}, including a wide range of quantum applications, circuit structures, and resource demands. All benchmarks are provided in OpenQASM format~\cite{cross2022openqasm} represented in the Clifford+1Q rotation gate set. We synthesize them into gate layers (GBC) or PPM layers (PBC) depending on the computing scheme using the tools adapted from~\cite{GridSynth, wang2024optimizing, wang2025tableau}. These benchmark include:

\vspace{2pt}
\noindent\textit{Arithmetic:} Adder-118, Adder-64, Multiplier-45, Multiplier-75~\cite{wang2025comprehensive}. Core to modular arithmetic in quantum algorithms such as Shor’s algorithm.

\noindent\textit{Quantum Simulation:} Ising-34, Ising-66, Ising-98~\cite{daley2022practical}. Simulate physical systems with practical relevance.

\noindent\textit{Quantum Fourier Transform:} QFT-15, QFT-29, QFT-63~\cite{coppersmith2002approximate}. Canonical quantum primitive with dense qubit interactions.

\noindent\textit{Variational Algorithms:} QAOA-30, VQE-SU2-24, VQE-2Loc-24~\cite{cerezo2021variational}. Hybrid quantum-classical workloads of interest for early FT systems and near-term advantage.

\vspace{2pt}
\noindent\textbf{Evaluation workflow.} We implement an end-to-end toolchain that converts benchmark circuits and produces FT workloads and reports end-to-end resources with per-component breakdowns under configurable FTQC/architectural settings
(e.g., GBC/PBC, code distances, \#MSF copies, NA--SC link latency, buffer size).
This enables rapid design sweeps for Sec.\ref{subsec: overall performance} and Sec.\ref{subsec: ablation}. In parallel, we
profile workload features (e.g., $T$-layer ratio, Pauli-weight timelines, active-set
deltas), which we use to explain HQA performance differences and derive
application-to-architecture rules.

\begin{figure*}[!ht]
    \centering
    \includegraphics[width=0.99\textwidth]{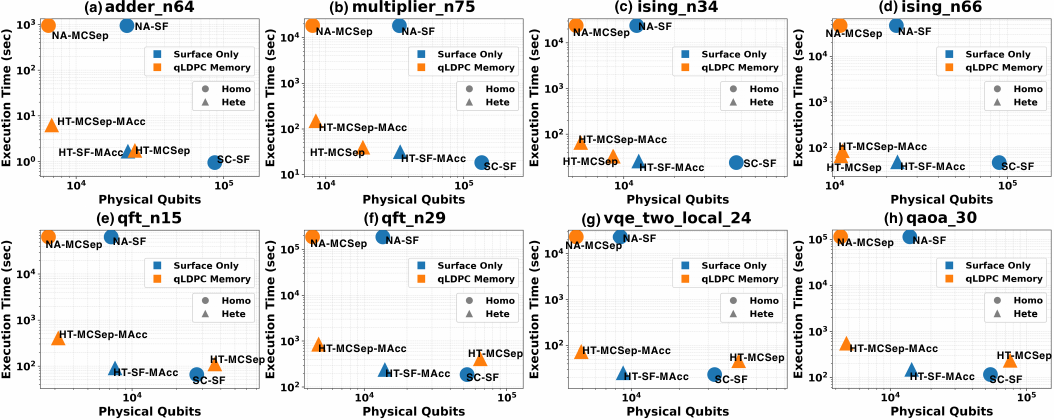} 
    \caption{
    Overall performance comparison across baseline architectures. For each benchmark, we plot execution time (y-axis) against physical qubit count (x-axis) on a log-log scale, with each point representing a different baseline. We selected 8 benchmarks out of 13 to present due to space limit.
    }
    \label{fig:overall-performance}
\end{figure*}

\subsection{Overall Resource Estimation} \label{subsec: overall performance}

We first evaluate the space/time overhead of architecture schemes within and across the two baseline groups (Table~\ref{tab:baseline-summary}), then discuss which designs offer the best space–time trade-offs across benchmarks and highlight the resulting design insights. The space-time overhead is presented in Fig.~\ref{fig:overall-performance}.

\noindent\textbf{Group 1: Surface-Code-Only (Blue markers).} 

\noindent\textbf{Space Overhead.}
\textbf{SC-SF} exhibits the largest space overhead—3.91× higher than \textbf{NA-SF}, primarily due to the need of routing ancilla path on SC platforms for PBC (Sec.~\ref{subsec:arch_compute}). In contrast, \textbf{NA-SF} benefits from GBC with transversal Clifford gates without routing ancilla, giving the most compact layout. \textbf{HT-SF-MAcc} introduces overhead over NA-SF due to buffer qubits for cross-platform MS delivery, but it remains mild (3\%) compared to the total qubit resource.

\noindent\textbf{Time Overhead.} Execution time is bounded by \textbf{NA-SF} on the high end and \textbf{SC-SF} on the low end, reflecting their 1000$\times$ clock speeds gap. \textbf{HT-SF-MAcc} sits between these extremes, achieving a \textbf{752$\times$} speedup over NA-SF and only \textbf{1.32$\times$} slowdown relative to SC-SF (the fastest execution time) on average. This \textit{massive speedup} mainly stems from the MSF-Clifford resource gap and MSF offloading to fast SC hardware, corresponding to $\rho_{\text{MS}}$ and $S$ as analyzed in Sec.\ref{subsec:MagicAcc analysis}. The \textit{slowdown} vs. SC-SF is due to executing Clifford layers on slower NA, but is mild because NA needs only one SM round per layer. Additionally, the realized speedup varies by benchmark according to its T-layer ratio $r_{T}=N_T/D$. Arithmetic workloads (adder, multiplier) with lower $r_{T}$ ($60\%$–$70\%$) yield more modest $500$–$600\times$ speedups; QFT benchmarks with higher $r_{T\text{layer}}$ ($90\%$) reach $700$–$800\times$ speedups; and Ising and VQE with nearly $100\%$ $r_{T}$ achieve up to $900$–$1000\times$, reflecting the Amdahl-style analysis in Sec.\ref{subsec:MagicAcc analysis}). 

We note that the large speedup stems from choosing the time-intensive MSC protocol for MSF; pairing MAcc with more time-efficient schemes such as MSD~\cite{litinski2019magic} would reduce the gain (smaller $\rho_{\text{MS}}$). However, in the near term, qubits—not wall-clock time—are the tighter constraint, making MSC the more practical choice over space-intensive MSD, especially with MAcc design. More broadly, when resource-state generation is time-heavy but space-efficient, MAcc offers a favorable space–time trade-off through acceleration.

\noindent\textbf{Group 2: qLDPC Memory (Orange markers).} We compare space–time overheads within Group 2 designs and against their Group 1 counterparts.

\noindent\textbf{Space Overhead.} \textit{Within Group 2,} \textbf{HT-MCSep-MAcc} (NA compute) incurs a mild 7.1\% space overhead compared to the most compact design \textbf{NA-MCSep}; the delta comes from buffer qubits reserved for magic-state delivery. In contrast, \textbf{HT-MCSep} (SC compute) uses 4.5$\times$ more qubits than HT-MCSep-MAcc—even though both employ space-efficient qLDPC memory. There are two key reasons: (i) SC compute requires wide ancilla path for PPM routing, inducing roughly 4$\times$ relative to an NA compute of equal size; (2) HT-MCSep must provision compute for the worst-case PPM weight, which is costly for benchmarks with heavy Pauli layers (e.g., QFT, VQE, QAOA; Fig.~\ref{fig:overall-performance}(e–h)). By contrast, HT-MCSep-MAcc can keep a smaller NA compute region and pay the difference in time (discussed later).

\noindent\textit{Across groups,} Group 2 schemes generally have lower space overhead than Group 1 (seen as the leftward shift of orange vs. blue markers), owing to qLDPC’s better space efficiency than surface codes. Crucially, the savings are workload-dependent: benchmarks with low active-qubit count $\{q_{\text{act}}^i\}$ (e.g., arithmetic) or low Pauli weight $\{w_{\text{Pauli}}^i\}$ (e.g., Ising, arithmetic, Fig.~\ref{fig:overall-performance}(a-d)) see up to 10.8$\times$ smaller footprints. Workloads with large Pauli weights and high memory-compute interchange (e.g., QFT, QAOA, VQE, Fig.~\ref{fig:overall-performance}(e-h)) benefit less—and can even exceed all-surface-code footprints—because they require larger compute patches and buffers.

\noindent\textbf{Time Overhead.} \textit{Within Group 2}, \textbf{HT-MCSep-MAcc} achieves a \textbf{237$\times$} speedup over NA-MCSep (Homo NA) and incurs a \textbf{2.25$\times$} slowdown relative to HT-MCSep. This mirrors the pattern in Group 1, where the T-layer ratio $r_{T}$ influences the magnitude of speedup (100$\times$-700$\times$) and slowdown (1.1$\times$-3.8$\times$). \textit{Across groups}, however, the speedup (slowdown) of Group 2 is more modest (pronounced) than their Group 1 counterparts (237$\times$, 2.25$\times$ vs. 752$\times$, 1.32$\times$). Two key factors contribute to this by introducing extra latency:
(1) Limited compute bandwidth. When the number of active qubits exceeds the available compute region, execution must be serialized across multiple steps;
(2) Store/load latency. When the active qubit difference between consecutive layers (gate or PPM) exceeds the logical buffer size, additional $d_{\text{qLDPC}}$ SM rounds are needed for each memory–compute interchange (see Sec.~\ref{subsec:arch_mcsep}).

\begin{figure*}[!ht]
    \centering
    \includegraphics[width=0.99\textwidth]{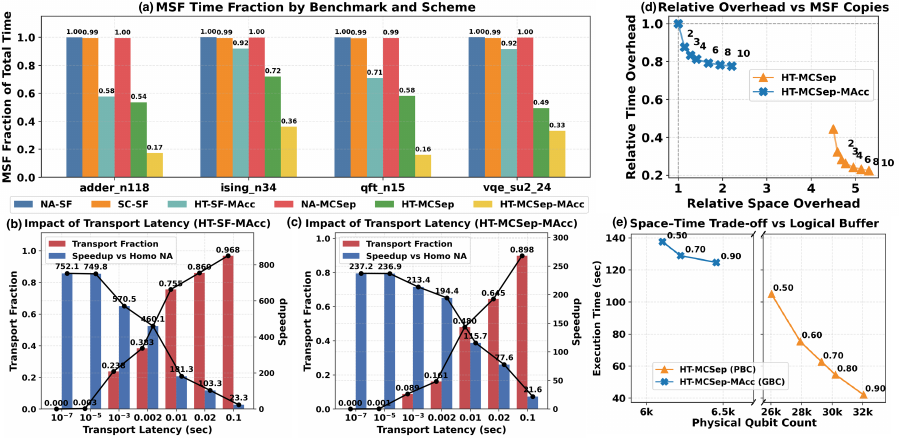} 
    \vspace{-3pt}
    \caption{
    Design space exploration: (a) Architecture designs can shift the latency contribution of MSF. MAcc largely reduces the MSF component. (b)(c) Transport latency component of the total execution time, which remains a small portion when it is around $10^{-3}s$ and yiels hundreds of speedup. (d)(e) Impact of MSF copies and logical buffer size on overall performance.
    }
    \label{fig:ablation}
     \vspace{-3pt}
\end{figure*}

\noindent\textbf{Space-Time Trade-off, Design Implications.} 
In summary, heterogeneous architectures (HT-SF-MAcc in Group 1 and HT-MCSep, HT-MCSep-MAcc in Group 2) offer highly favorable space–time trade-offs compared to their homogeneous counterparts, approaching the minimal execution time of Homo SC and the compact qubit footprint of Homo NA. This advantage arises from combining the strengths of both platforms by assigning them best suited roles: fast magic state generation on SC hardware, paired with dense qLDPC memories or space-efficient compute on NA. 

\textit{However, the optimal design choice varies by applications.} Benchmarks with low active qubit counts benefit more from \textbf{HT-MCSep-MAcc} with NA compute, while those with low Pauli weights in PPM layers favor \textbf{HT-MCSep} with SC compute. This insight is particularly valuable when selecting architectures for applications with contrasting characteristics—e.g., \textbf{QFT} (low active qubits, high Pauli weights) favors HT-MCSep-MAcc, while \textbf{Ising} (high active qubits, low Pauli weights) favors HT-MCSep—for achieving the best space–time efficiency. Overall speaking, HT-MCSep-MAcc emerges the most promising near-term option among these baselines, typically requiring $<10^{4}$ physical qubits and running in $10^3s$ across these benchmarks.



\subsection{Design Space Exploration and Analysis} \label{subsec: ablation}
We perform component-wise and design-space analyses to isolate the impact of key factors: (i) MSF latency, (ii) cross-modality transport latency, (iii) MSF parallelism (number of copies), and (iv) logical-buffer capacity. For each, we quantify effects on total runtime, attainable speedup, and shifts along the space–time frontier.

\noindent\textit{Why this analysis is feasible.}
These what-if studies are enabled by a unified workflow that turns FT workloads into parameterized cost traces (compute, MSF, transport, store/load) and replays them under new configurations in seconds. This lets us vary one factor at a time—without re-synthesizing or re-encoding—and attribute changes in runtime, speedup, and space–time position to that factor alone. Prior evaluations typically lacked such cross-modality, end-to-end reconfigurability, making rapid partitioning studies impractical.

\noindent\textbf{1. MSF Latency Fraction.} As shown in Fig.~\ref{fig:ablation}(a), MSF dominates runtime in all homogeneous baselines (NA-SF, SC-SF, NA-MCSep), consuming nearly 100\% of the total latency. In contrast, heterogeneous architectures significantly reduce the MSF latency fraction by accelerating MSF on SC hardware, resulting in a more balanced latency distribution across operations. Among heterogeneous baselines, memory–compute separation (HT-MCSep, HT-MCSep-MAcc) has lower MSF dominance due to the additional store/load latency compared to HT-SF-MAcc, matched with analysis in Sec.\ref{subsec:MCSep analysis}. Between them, HT-MCSep-MAcc (NA compute) exhibits a lower MSF fraction than HT-MCSep (SC compute), since slower Clifford gates on NA make MSF relatively less significant.

\noindent\textbf{2. Impact of Transport Latency.} We vary the interconnect latency from $10^{-7}$s to $10^{-1}$s for HT-SF-MAcc and HT-MCSep-MAcc, each using a single MSF copy with the cultivation protocol. Fig.~\ref{fig:ablation}(b,c) shows that transport latency becomes increasingly impactful, with its contribution reaching $23.8\%$ and $8.9\%$ of total runtime at $10^{-3}$s, respectively. Nevertheless, both schemes still deliver substantial speedups ($571\times$ and $213\times$) over their homogeneous NA baselines. This reveals the tolerable range of transport latency within which heterogeneous architectures retain superiority.


\noindent\textbf{3. Impact of MSF Copies.} We evaluate the performance–cost trade-off by increasing the number of MSF copies in HT-MCSep and HT-MCSep-MAcc. Fig.\ref{fig:ablation}(d) shows that additional MSFs reduce execution time but with diminishing returns and growing space overhead. Notably, using 3 MSFs strikes a good balance between performance and cost. Crucially, increasing MSFs alone in HT-MCSep-MAcc (NA compute) cannot surpass the execution time of HT-MCSep (SC compute), due to the persistent Clifford latency bottleneck on NA.

\noindent\textbf{4. Impact of Logical Buffer.} We sweep the quantile parameter that determines the logical buffer size (Sec.~\ref{subsec:arch_mcsep}) and evaluate its effect on space–time trade-off. As Fig.\ref{fig:ablation}(e) shows, larger quantiles increase logical buffer region and thus space overhead, but reduce memory stall time. Notably, Fig.\ref{fig:ablation}(e) also suggests that quantile tuning has higher impact on PBC (HT-MCSep) than GBC (HT-MCSep-MAcc). This is because inter-PPM-layer Pauli-weight deltas vary more widely than inter-gate-layer active-qubit deltas, so buffer sizing has a larger impact under PBC. Consequently, GBC is less sensitive to buffer scaling.

\section{Conclusion}
This work represents an initial step toward HQA design. We explored two role-assignment strategies and demonstrated how complementary hardware properties can be leveraged to reduce fault-tolerant overheads. While our results highlight clear potential, we emphasize that this study is far from exhaustive or complete. Other role partitions may prove equally or more effective, and more detailed optimizations will require full-stack integration across compilers, runtimes, and hardware to really fulfill this potential. We view our analysis as a starting point rather than a definitive solution, and hope that it contributes to shaping a broader research agenda for heterogeneous quantum computing. 

\bibliographystyle{IEEEtranS}
\bibliography{References}

\end{document}